# Dynamic Spin Fluctuations in the Frustrated *A*-site Spinel CuAl$_2$O$_4$

Hwanbeom Cho[1,2,3*], R. Nirmala[4], Jaehong Jeong[2,3], Peter J. Baker[5], Hikaru Takeda[6], Nobuyoshi Mera[6], Stephen J. Blundell[1], Masashi Takigawa[6], D. T. Adroja[5,7], and Je-Geun Park[2,3,8*]

[1]Clarendon Laboratory & Physics Department, University of Oxford, Parks Road, Oxford, OX1 3PU, United Kingdom
[2]Department of Physics & Astronomy, Seoul National University, Seoul 08826, Korea
[3]Center for Correlated Electron Systems, Institute for Basic Science, Seoul 08826, Korea
[4]Department of Physics, Indian Institute of Technology Madras, Chennai 600 036, India
[5]ISIS Facility, STFC Rutherford Appleton Laboratory, Harwell Oxford, Oxfordshire OX11 0QX, United Kingdom
[6]Institute for Solid State Physics (ISSP), University of Tokyo, Kashiwa, Chiba 277-8581, Japan
[7]Highly Correlated Matter Research Group, Physics Department, University of Johannesburg, P.O. Box 524, Auckland Park 2006, South Africa.
[8]Center for Quantum Materials, Institute for Basic Science, Seoul 08826, Korea

*Corresponding Author: hwanbeom.cho@physics.ox.ac.uk & jgpark10@snu.ac.kr

**ABSTRACT**

We performed nuclear magnetic resonance (NMR) and muon spin relaxation ($\mu$SR) experiments to identify the magnetic ground state of the frustrated quantum *A*-site spinel, CuAl$_2$O$_4$. Our results verify that the ground state does not exhibit a long-range magnetic ordering, but a glass-like transition manifests at $T^*$=2.3 K. However, the Gaussian shape and the weak longitudinal field dependence of $\mu$SR spectra below $T^*$ show that the ground state has dynamic spin fluctuations, distinct from those of conventional spin-glasses.

## I. INTRODUCTION

The *A*-site spinel system, where magnetic ions occupying the crystallographic *A*-site form a diamond lattice, has been extensively studied over the last decade or so as a candidate for the novel frustrated magnetic system. When it is influenced under antiferromagnetic nearest ($J_1$) and next-nearest ($J_2$) neighbor exchange interactions, its magnetic ground state can have noncollinear magnetic structures. Depending on the strength of the quantum fluctuations [1] and the ratio of the exchange interactions ($J_2/J_1$) [2], the ground state of the *A*-site spinel shows a complex phase diagram. For instance, when $S\leq1$ and $J_2/J_1>1/8$, there is no thermal phase transition, and the ground state has the superposed state of spiral states with propagation vectors ***q*** spanning a particular surface in momentum space [2]. For $S>3/2$, a quantum-to-classical crossover occurs, followed by a thermal phase transition into an ordered ground state [1]. Other than the spin size and the ratio of the exchange interactions, however there is an additional factor to be considered for spinel compounds. It is a site-disorder $\eta$ due to the fractional occupancy of atoms in the tetrahedral $A$- and octahedral $B$-sites. It can be another critical parameter affecting the ground state nature of the frustrated $A$-site spinel.

While there seems to be less clear theoretical understanding on the non-zero disorders effect in the $J_1$-$J_2$ spinel system [3], many experimental studies suggest a ground state that is distinct from a conventional long-range ordered one [4-7]. For example, $MnAl_2O_4$ ($S$=5/2) orders collinearly with prominent spin fluctuations, while $FeAl_2O_4$ ($S$=2) shows a spin freezing transition with a memory effect which is distinct from canonical spin-glass behavior [4]. In the case of $CoAl_2O_4$ ($S$=3/2), the ground state has a collinear magnetic ordering with a short-range dynamic order when $\eta$~6 % [5]. There are also reports that $CoAl_2O_4$ has a spin-liquid phase or coexisting phases of spin-liquid and glass for a particular range of the disorder [6,7]. By comparing the ground states of $FeAl_2O_4$, $MnAl_2O_4$, and $CoAl_2O_4$, we can observe that the correlated states change from a long-range to a short-range order as the spin size reduces. This tendency motivated us to explore an $A$-site spinel system across the quantum crossover ($S\leq1$) to search for quantum effects in the frustrated system.

$CuAl_2O_4$ is a spinel compound, where magnetic $Cu^{2+}$ and nonmagnetic $Al^{3+}$ ions mainly occupy the $A$- and $B$-sites, respectively (Fig. 1a). There is a finite amount of disorder: according to our high-resolution x-ray and neutron diffraction studies, about 30 % of the $A$-sites (and 15 % of the $B$-sites) are occupied by $Al^{3+}$ ($Cu^{2+}$) ions [8]. We tried various synthesis methods and heat treatments to reduce the disorder, but this level of site-mixing is seen to be present in all our samples. Intriguingly, $CuAl_2O_4$ has a cubic lattice of $Cu^{2+}$ with spin-half, although $Cu^{2+}$ ions sit at the center of a crystal electric field made by surrounding four oxygen ions forming a tetrahedron cage and are subjected to the Jahn-Teller effect. The cubic lattice of $CuAl_2O_4$ has recently been proposed to host the emergence of a spin-orbital-entangled $J_{eff}$=1/2 state instead of a spin-only $S$=1/2 state [9]. We should note that according to the theoretical studies, the spin-orbit entangled state can survive even with 50 % of the site-disorder.

Another interesting point is that the magnitude of exchange interactions of the system estimated by its Curie-Weiss temperature ($\theta_{CW}$=−140 K) is quite large. Nonetheless, the ground state shows no long-ranged order down to 0.4 K, but a correlated spin state instead emerge at $T^*$=2.3 K according to both neutron powder diffraction and heat capacity measurements [8]. The frustration factor ($|\theta_{CW}|/T^*$~60) thus obtained, a standard measure of magnetic frustration, puts $CuAl_2O_4$ as one of the most frustrated systems among all the $TAl_2O_4$ spinels ($T$=Mn, Fe, Co, Ni, Cu) [10,11]. This aspect implies that sharp spin fluctuations of the quantum spin might dominate the ground state of $CuAl_2O_4$.

Besides, the heat capacity of $CuAl_2O_4$ shows a nonlinear temperature dependence, which is also distinct from a canonical spin-glass state [8]. The enhanced quantum fluctuations estimated from the small size of the magnetic moments and the significant frustration factor are expected to produce nontrivial dynamics in the ground state of this $A$-site spinel compound. In this paper, using local probes of magnetism, namely, nuclear magnetic resonance (NMR) and muon spin relaxation ($\mu$SR), we report that the ground state of CuAl$_2$O$_4$ does indeed not order or freeze completely. Still, it has significant spin fluctuations, which persist well below $T^*$.

**II. METHODS**

High quality polycrystalline CuAl$_2$O$_4$ samples were synthesized by a solid-state reaction method. Initial raw materials of CuO (99.995 %, Alfa Aesar) and Al$_2$O$_3$ (99.995 %, Alfa Aesar) were mixed in a 1:1 molar ratio, and the mixture was pelletized to be sintered at 1000 ºC in the air for 2-3 days with several intermediate grinding steps. Single-crystals of CuAl$_2$O$_4$ were grown using a flux method [12]. CuO and Al$_2$O$_3$ were mixed homogeneously in a 1:1 molar ratio with a sufficient amount of flux material, which is anhydrous sodium tetraborate (99.998 %, Alfa Aesar). The mixture was contained in a platinum crucible and melted using a ceramic tube furnace flowing O$_2$ gas. After heating up to 1350 ºC, it was dwelled for 24 hours to make the batch molten homogeneously and then slowly (1 ºC/h) cooled down to 750 ºC. After removing the flux by applying diluted HCl (~17 %), we separated dark brown single-crystals of CuAl$_2$O$_4$ (photo shown as an inset in Fig. 1b) from the flux. The crystal size obtained measures up to 1 mm$^3$ with typical octahedral morphology of spinel-type crystals. The purity of the polycrystalline and single-crystal samples was further checked by powder and single-crystal x-ray diffraction (XRD), respectively.

High-resolution powder diffraction experiment was performed (Fig. 1c) using a commercial powder diffractometer, D8 Discover (Bruker), with wavelengths $\lambda_{K\alpha}$ of 1.540590 and 1.544310 Å ($K_{\alpha 1}$ and $K_{\alpha 2}$, respectively). XtaLAB P200 (Rigaku) with Mo source ($\lambda_{K\alpha}$=0.710747 Å) was used to identify the structure of the single-crystals. The number of total independent reflections given by the single-crystal diffraction experiment was 148 (Fig. 1b). The observed intensity, $F^2_{obs}$, and the calculated intensity, or square of structural factor, $F^2_{calc}$ of each reflection points are well-matched with each other, and this leads to reasonably good agreement factors: $R_{F2}$, $R_{wF2}$, and $R_F$. The whole analysis was done through the integrated intensity refinement method powered by *FullProf* software [13].

We measured the bulk DC magnetic susceptibility of the CuAl$_2$O$_4$ single-crystal using a commercial equipment (MPMS-3, Quantum Design). We applied an external magnetic field parallel to the crystallographic [0 0 1] direction of the crystal. The NMR and $\mu$SR experiments were conducted to identify the characteristics of the local moments. We measured the NMR spectra of a single-crystal CuAl$_2$O$_4$ under a constant external magnetic field of $B_0$=20 kG, scanning around the Larmor frequency of $^{27}$Al with a nuclear spin $I$=5/2. $\mu$SR experiments were done at the ISIS muon source (MuSR) with a polycrystalline sample. We identified the temperature and field dependence of the muon depolarization spectra by cooling down to 1.4 K and by changing the longitudinal magnetic field up to 3 kG. The $\mu$SR data [14] with high statistics, obtained for 0.1<$t$<32.29 μs, were analyzed using a $\mu$SR analysis package, *Mantid* [15].

Density functional theory (DFT) calculations were carried out using the plane-wave program QUANTUM ESPRESSO [16] and utilized the generalized gradient approximation for the exchange-correlation functional [17]. The ions were modeled with ultrasoft pseudopotentials [18], while a norm-conserving hydrogen pseudopotential was used to model the muon. The energy cutoffs for the wave function and the charge density were set to 60 and 600 Ry, respectively, and the integration over the Brillouin zone was carried out using a 3×3×3 Monkhorst-Pack $k$-space grid [19].

## III. RESULTS
### III-A. Crystal structure
$CuAl_2O_4$ has a cubic spinel lattice with space group $Fd\bar{3}m$ (#227) and the lattice parameter $a$=8.083 Å (see Fig. 1 and Table 1). No evidence of tetragonal distortion is observable within the resolution of our high-resolution powder and single-crystal diffractometers, even though $Cu^{2+}$ is a Jahn-Teller active ion. The $A$-sites of magnetic $Cu^{2+}$ ions form a diamond lattice while the $B$-sites of nonmagnetic $Al^{3+}$ ions have a pyrochlore structure. There is the site-mixing between $Cu^{2+}$ and $Al^{3+}$ ions ($[Cu_{1-\eta}Al_\eta]_A[Al_{2-\eta}Cu_\eta]_BO_4$, $\eta$=0.3), which can lead to disordered magnetic interactions in this system. Other detailed structural information obtained from the refinement, such as the fractional atomic positions and thermal parameters, is summarized in Table 1.

### III-B. NMR measurement
The observed NMR spectra show the central peak around 22.19 MHz (Fig. 2a), which corresponds to the Larmor frequency of $^{27}Al$, $\omega_0=\gamma B_0$=22.1929 MHz, with the gyromagnetic ratio of $^{27}Al$, $\gamma$=1.109406 MHz/kG and the external magnetic field $B_0$=20 kG. As the temperature decreases, the width of the spectra gets broadened. It is related to the enhancement of internal fields upon the reduction of thermal fluctuations and the onset of a correlated state. In particular, we could not observe any evidence of long-range magnetic ordering, such as peak splitting below $T^*$ [20].

According to the $^{27}Al$ MAS (magic-angle spinning) NMR spectrum of a spinel analog $MgAl_2O_4$ [21,22], the asymmetric central peak close to $\omega_0$ of $CuAl_2O_4$ originates from Al in the $B$-site. Since the peak shifts within 0.04 % down to the base temperature, the information obtained via the central peak can be related to the internal fields in the $B$-site for the whole temperature range. The temperature dependence of the inverse of the full width at half maximum (FHWM) of the NMR spectra matches well with the inverse susceptibility (Fig. 2b). Since the width of NMR spectrum is related to the distribution of internal fields on the $B$-site and the field distribution is associated with the disorder between $Cu^{2+}$ and $Al^{3+}$ ions [23], the correspondence between the spectral width and the bulk susceptibility indicates that the bulk susceptibility originates mainly from magnetism induced by the disorder.

Meanwhile, the peak shift, $K$ (%) from the resonance frequency $w_0$ is another quantity that directly probes the local moments free from defects [20,23]. The magnitude of $K$ increases up to $T$=4.2 K and then drops down at $T$=1.4 K as the temperature decreases, indicating a change in the magnetic ground state across $T^*$. A noticeable difference between the field-cooled (FC) and the zero-field-cooled (ZFC) spectra is seen to develop as the temperature goes down below $T^*$ (Figs. 2c and 2d). It shows that the correlated ground state behaves somewhat like a spin-glass state.

In the case of $^{27}$Al with $I=5/2$, the Zeeman splitting and the first-order nuclear quadrupolar interactions [24] produce four additional satellite peaks, coming from the transitions between $I_z=\pm 1/2$ to $\pm 3/2$ and $\pm 3/2$ to $\pm 5/2$ states, around the central peak at $\omega_0$ (transition between $I_z=\pm 1/2$). The NMR spectra of other spinel compound $CoAl_2O_4$ shows a five-peak structure, and the detailed feature changes depending on the direction of the external magnetic field $B_0$ [25]. The minimum width of the spectra coincides with the angular factor of the quadrupolar interactions $(3\cos^2\theta-1)/2$ (where $\theta$ is the angle between $B_0$ and the three-fold rotational symmetry ($C_3$) axis of the $B$-site) becoming zero, which occurs, $B_0 \parallel [0\ 0\ 1]$.

However, we barely observed such an angular dependence in our $CuAl_2O_4$ spectra. Figs. 2a and 3 show the spectra taken at 200 K of a sharp central and broad satellite peaks. The satellite peaks are too broad to be resolved into a four-peak structure in our spectrum. And the intensity of the satellite peaks does not vary visibly when the orientation of the crystal is changed. Moreover, there are still broad shoulder peaks, which should disappear when the field is parallel to [0 0 1]. These characteristics can be understood as the effect of site-disorder. NMR spectra are sensitive to local fields surrounding the Al-site, and the NMR peaks can become broadened when the local fields are deformed. As the Al-site forms a pyrochlore structure, there are six neighboring $Al^{3+}$ ions around one $Al^{3+}$ ion, where some of them are interchanged with $Cu^{2+}$ ions. It naturally produces a distribution of local fields, which broadens the peaks [26]. Since the local fields have no preferred orientation (as for a powder sample), then angular-independent broadening emerges in both the primary and satellite peaks.

### III-C. $\mu$SR measurement
Muon spin relaxation ($\mu$SR) is another essential tool for the study of local magnetism. In particular, it can identify fast spin dynamics well outside of the detection window of the NMR technique [27]. Also, it is possible to study a ground state without applying an external magnetic field, which may modify the nature of a ground state. The $\mu$SR spectra can be normalized following a relation $A(t)=A_1P(t)+A_c$, where $A_1$ and $A_c$ are a normalization factor and a constant background, respectively. Above 3 K, the depolarization spectra $P(t)$ shows Gaussian decay (Fig. 4a), which can be fitted to the following decay function:

$$(1) \quad P(t) = [\tfrac{1}{3} + \tfrac{2}{3}(1-(\Delta t)^2)e^{-(\Delta t)^2/2}]e^{-\lambda t},$$

where the term inside the square bracket indicates a static Gaussian Kubo-Toyabe function [28] of the nuclear dipole moments with a distribution of $\Delta$. The exponential term denotes fast paramagnetic fluctuations of the electron moments of $Cu^{2+}$ ions with the relaxation rate of $\lambda = 2\Delta_e^2\nu/(\nu^2+\gamma_\mu^2(\mu_0 H_{LF})^2)$ [29,30], where $\Delta_e$ is the width of a distribution of local fields induced by the electrons, $\nu$ is the fluctuation frequency of the local field at the muon site, $\gamma_\mu$ is the gyromagnetic ratio of muon (85.1 MHz/kG), and $H_{LF}$ is the applied longitudinal magnetic field (equals to zero for the zero-field data). In the high-temperature region (6-15 K), paramagnetic fluctuations are too fast ($e^{-\lambda t} \approx 1$). And the nuclear contribution dominates over the depolarization with a constant value of $\Delta \cong 0.2$ MHz (Fig. 4b), which is similar to the nuclear field distribution of another Cu-based spinel, $CuGa_2O_4$ [31]. As the temperature decreases, however, the correlation of electron moments starts to develop, and it forms finite local internal fields that can be decoupled from the nuclear moments. This picture corresponds well with the measured temperature dependence of $\Delta(T)$, which starts to reduce below 5 K and finally vanishes at 3 K. Moreover, $\lambda(T) \propto 1/\nu$ gets enhanced almost by two orders of magnitude as the paramagnetic fluctuations reduce.

Below $T=3$ K and down to the base temperature 1.4 K, we could not observe an oscillating feature that is expected for a long-ranged magnetic order. We did several trials to fit the spectra using non-oscillating models, such as single stretched exponential and two exponential functions. Still, we failed to get a proper fitting with realistic physical parameters. The zero-field spectra could only satisfactorily be fitted to an empirical decay function,

$$(2) \quad P(t) = xe^{-(\lambda_1 t)^\beta} + (1-x)e^{-\lambda_2 t},$$

which is composed of two different decay channels. The result of the fitting is shown as solid lines in Fig. 4a. As the temperature goes down, the portion of the first term $x$ increases and dominates at the base temperature (Fig. 5c) (this point is discussed later), with the exponent $\beta$ leveling off to 2 from 1 (Fig. 4b). Such initial Gaussian decay below $T^*$ is different from conventional spin-glasses in the narrowing limit ($\nu \gg 5\Delta_e$), where the polarization decays exponentially ($\beta=1$) for a dense system or root-exponentially ($\beta=1/2$) for a diluted one [32]. In addition to the first term in Eq. (2), which fits the primary feature of the relaxation spectrum at the base temperature, there is an additional exponential term necessary to fit the tail of the $\mu$SR spectra. The decay rates $\lambda_1$ and $\lambda_2$ seem to level off to finite values at 1.4 K (Fig. 4b), which is the characteristic behavior of a dynamic ground state with significant quantum fluctuations [32,33]. These decay rates are comparable to what we obtained from $1/T_1$, of which $T_1$ is the relaxation time to depolarize down to $1/e$, and it does not depend on a model of a decay function used. In the case of a slowly fluctuating limit, however, the relaxation rate is rewritten to $\lambda = 2\nu/3$ [30], and it decreases as temperature goes down. It makes a sharp peak at the transition temperature for a statically freezing state [34], which is different from our result.

We further measured the muon spectra with applied longitudinal magnetic field $H_{LF}$ to elucidate the dynamics of the ground state [35]. In the case the spin fluctuations in a ground state are significant, compared with the static case, a larger $H_{LF}$ is needed to cause the decoupling of the muon spin from local moments. Fig. 5a shows the asymmetry below $T^*$ under various $H_{LF}$. If the magnetic ground state of $CuAl_2O_4$ is frozen, a longitudinal field $\mu_0 H_{LF} = 10\Delta_e/\gamma_\mu \approx 10 \cdot (1/T_1)/\gamma_\mu \approx 320$ G will be sufficient to decouple the muon spins [33,35]. However, the moments of $CuAl_2O_4$ cannot be entirely decoupled even up to 3000 G, which provides a clear evidence for a dynamic ground state. To analyze the data quantitatively (Fig. 5b), we firstly tried fitting the data using the dynamic Kubo-Toyabe (DKT) decay function [28]. For $\mu_0 H_{LF} = 30$ G, which is large enough only to decouple the nuclear moment, the depolarization curve can be fitted with $\Delta_e$=3.28(3) μs$^{-1}$ and $v$=2.64(6) μs$^{-1}$ (line 1). However, using this model with these fitting parameters, the calculated polarization should be decoupled entirely under fields above $\mu_0 H_{LF}$=1000 G (line 2). The conventional DKT model, therefore, cannot be adopted to demonstrate the observed depolarization spectra. To fit the data under the longitudinal fields, we used the decay function from Bono *et al.* [32,36],

$$(3) \quad P(t) = x G_z^G(t, f\Delta_e, f H_{LF}, fv) + (1-x) e^{-\lambda_2' t},$$

where $G_z^G$ is the DKT decay function with a scaling factor *f*. The first term originates from the model with 'sporadic field fluctuations,' where the muon can observe local moments only in the fraction *f* of the total relaxation time *t* and the dynamics of the state are not affected by a longitudinal field [32,36]. This model was introduced first to explain frustrated kagomé bilayer compounds $SrCr_{9p}Ga_{12-9p}O_{19}$ (SCGO) and $Ba_2Sn_2ZnGa_{10-7p}Cr_{7p}O_{22}$ (BSZGCO), where a singlet state occasionally breaks into propagating unpaired spins (spinons). At the same time, the local moments of the systems can be detected during the short time *ft* when the unpaired spins emerge from the dimer [32,36]. For the rest of the experimental time, the polarization relaxes following the exponentially decaying function (the second term in Eq. (3)) representing other electron spins, which fluctuate according to a conventional Markovian process [28,30]. Using this model, we can successfully reproduce the whole field-dependent data (solid lines in Figs. 5a and 5b) with the fitting parameters of $f \approx 0.046$, $\Delta_e \approx 117$ μs$^{-1}$, and $v \approx 115$ μs$^{-1}$. These parameters are comparable to other systems that adopt similar models [32,36,37]. Notably, the fluctuation rate is as fast as that of a quantum spin-liquid candidate herbertsmithite [38] and almost two orders of magnitude faster than the fluctuation rate, which can be detected by NMR [27]. The weight *x* saturates to the maximum under zero-field (inset of Fig. 5c), which indicates that the major spin dynamics of the ground state is demonstrated by the sporadic field fluctuation process. Moreover, the fact, (1- *x*)>0 denotes the coexistence of minor phase, having relatively slower spin fluctuations $v \approx 7$ μs$^{-1}$ [Appendix].

Now, we can understand that the Gaussian decay of muon depolarization by utilizing the dynamic model with the sporadic field fluctuations. According to the conventional Markovian modulation of internal fields, a rapidly fluctuating state shows exponential decay on the limit of $v \gg \Delta_e$ [28,30]. Despite its remarkable local field fluctuations ($v \approx 115$ μs$^{-1}$), depolarization of $CuAl_2O_4$ can show a Gaussian decay because its local fields are in a region of $v \leq \Delta_e$ [32,36,37]. Moreover, the sporadic emergence of the local fields leads to the relaxation of having an effectively reduced local field distributions $f\Delta_e$. It makes the 'actual' longitudinal field necessary to decouple the local moments (10·$\Delta_e/\gamma_\mu \approx$ 14 kG) much larger than the external field evaluated from the initial decay rate (10·(1/$T_1$)/$\gamma_\mu \approx$ 320 G).

In addition to the first term in Eqs. (2) and (3), which depicts the dynamic nature of the ground state, there is an additional exponential term necessary to fit the tail of the μSR spectra. Like some of the spin-glasses showing exponential decay with moderate spin fluctuations [39,40], the other term can be related to the glassy phenomena observed in the previous bulk magnetic measurements [8] and our NMR results. Therefore, the ground state of $CuAl_2O_4$ is composed of a major (*x*=0.7) dynamic phase with the sporadic spin fluctuations and a minor (1−x=0.3) phase, rendering its glassy nature. Interestingly, $x$ at the base temperature and under zero-field corresponds to the ratio of $Cu^{2+}$ ions in the $A$-site (1−$\eta$=0.7). The coexistence of dynamic and glassy states also was observed in kagomé systems SCGO and BSZGCO [32,36]. As the temperature (or $H_{LF}$) increases, $x$ decreases (Fig. 5c), which indicates that the dynamic state transforms into the state, in which spins fluctuate following the classical Markovian process.

To estimate the potential muon stopping sites in $CuAl_2O_4$, we employed DFT calculations to map out the electrostatic Coulomb potential within the unit cell. The maxima of such potential map correspond to likely locations for adding a positive charge, such as the muon [41,42]. This analysis leads to a single candidate muon site for 'ideal' $CuAl_2O_4$ (neglecting any site-mixing) with site symmetry 96$g$ and fractional coordinates of around [0.04, 0.04, 0.91]. This site is approximately 1.4 Å away from an oxygen anion and about 1.9 Å away from two copper cations. We also repeated the calculations with configurations, which took into account site-mixing. And there was a clear energetic preference for the muon to choose a location close to a copper cation sitting on an octahedral $B$-site (occupied by aluminum in the ideal case). The site in this case (shown in Fig. 6) is closest to two oxygen anions but is only ~2 Å separated from each three copper cations (the $B$-site one that is due to site-mixing and two $A$-site ones). One might then expect that the muon-spin relaxation is dominated by correlated fluctuations on this triangle of copper spins. For our experiment, where the site-mixing is quite significant ($\eta$=0.3), we expect that muons will preferentially locate close to $B$-site copper anions, as in Fig. 6. Note that this site is only approximate, as the electrostatic potential approach does not consider local distortions caused by the muon itself. However, we can estimate the extent of the likely induced distortion based on experience with pyrochlore oxides [42], and this will not alter the conclusion that the muon will be dominated by fluctuations from a triangle of copper spins.

## IV. DISCUSSION

Our NMR spectra, at first glance, show several features related to a normal spin-glass state: the absence of a long-ranged order, the onset of metastability separating the ZFC and FC spectra, and the unresolvable satellite peaks. These are most likely to originate from the distribution of crystal electric fields and the randomized exchange interactions coming from the finite amount of disorder in the system. The excellent matching of the width of NMR spectra and the bulk susceptibility (Fig. 2b) indicates that the bulk magnetic properties relating to the glassy nature mainly reveal an effect of the disorder. Meanwhile, our $\mu$SR measurements, as a faster local probe to observe spin dynamics, also verified notable dynamics below $T^*$, which is not seen in a conventional spin-glass: the initial Gaussian-like decay, the saturation of the relaxation rate, and the much weaker $H_{LF}$ dependence than a frozen state.

We carefully examined the reports for other spinel analogs to understand further the effect of the site-disorder on the frustrated diamond lattice. For example, $[Co_{1-\eta}Al_\eta]_A[Al_{2-\eta}Co_\eta]_BO_4$, having a classical spin-3/2 in the diamond lattice, has an antiferromagnetic collinear ordering [5] for the minimum $\eta$ because the ratio $J_2/J_1=0.11$ [6] is below 1/8. When the disorder is introduced, the long-ranged order transforms into a short-ranged order to induce a spin-glass phase [7,43]. Notably, up to $\eta=0.36$, the critical temperature $T^*$, at which FC and ZFC magnetic susceptibility curves start to bifurcate, reduces as $\eta$ increases [7,43]. However, in the region of $0.36<\eta<0.77$, $T^*$ increases closely following the trend of $\eta$.

This different disorder-dependency can also be interpreted as follows. The magnetic susceptibility data [43] and the $\eta$-$T$ phase diagram [7] of the low-disorder system correspond to those of a diluted diamond lattice, such as $Co_{1-x}Zn_xAl_2O_4$ [44], where the number of Co ions in $A$-site is variable but that in the $B$-site is controlled. It indicates that the low-disorder system can be approximated as the diluted diamond system [45]. In the low-disorder region, the reduction of $T^*$ occurs because the nonmagnetic ion ($Zn^{2+}$), introduced in the $A$-site, hinders the magnetic interactions between Co ions in the $A$-site. On the other hand, the interactions between the $A$-sites becomes negligible in the high-disorder region. And the dominant interactions exist between the $A$- and $B$-sites, and the $B$-sites. The increased number of magnetic ions ($Co^{2+}$) in the $B$-site, which reduces the average distance between the magnetic atoms in the $B$-sites, eventually leads to the rise of $T^*$ [43].

Following the above example, our system $[Cu_{1-\eta}Al_\eta]_A[Al_{2-\eta}Cu_\eta]_BO_4$, $\eta=0.3$ can be understood as a diluted diamond lattice-like $[Cu_{1-\eta}Al_\eta]_A[Al_2]_BO_4$. The nature of the diluted diamond lattice with the quantum spin-1/2 was previously studied in $Cu_{1-\eta}Zn_\eta Rh_2O_4$ [46,47]. Contrary to the collinear ordering of cobalt spinel such as $CoAl_2O_4$ and $CoRh_2O_4$, the ground state of $CuRh_2O_4$ is reported to have an incommensurate helical order [46]. Its average $J_2/J_1=0.125$ meets the condition [1] of being proposed to induce the spiral spin-liquid phase without the thermal phase transition into a magnetically ordered ground state. However, as $CoRh_2O_4$ has a distorted tetragonal lattice, the exchange interactions become separated into the in- and out-of-plane terms, which produces the ordered ground state [1,46]. Moreover, the ordering of $Cu_{1-\eta}Zn_\eta Rh_2O_4$ becomes weaker [47], and fast spin fluctuations emerge as $\eta$ increases above 0.44. It is because that the slight local distortion induced by the doping leads to modulations of the interactions, stabilizing adjacent other helical order, and eventually it induces a competition between the phases.

From the similar Curie-Weiss temperature of $CuRh_2O_4$, $\theta_{CW}=-132$ K [46], we anticipate $CuAl_2O_4$ to be close to the Lifshitz point $J_2/J_1=1/8$ [3]. Unlike tetragonal $CuRh_2O_4$, the cubic lattice of $CuAl_2O_4$ [48] can, moreover, host rather isotropic $J_1$ and $J_2$ interactions, which is a favorable condition to induce the spiral liquid state. We note that the fast spin fluctuations observed in $CuAl_2O_4$ cannot be explained by the two previous reports. One is the ground state of $[Cu_{0.7}Zn_{0.3}]Rh_2O_4$ with an order at 9 K [47]. Another is on $[Co_{0.6}Zn_{0.4}]Al_2O_4$, where the heat capacity shows a nearly linear temperature dependence of a canonical spin-glass state [49] below $T^*\sim 5$ K [44]. The fluctuations are known to become more substantial when the 'stiffness' of the bulk spiral vanishes on approaching the Lifshitz point [3]. Therefore, we can ascribe the dynamic term of $CuAl_2O_4$ to the liquid state. The reduction of $x$ depending on the external magnetic field in the inset of Fig. 5c is then naturally related to breaking the degeneracy of the superposed manifold of the liquid state by applying the field. The site-disorder, moreover, can also lift the degeneracy of the state, inducing a finite region around it, in which the spins are deformed from an ideal spiral pattern [3]. It can then stabilize the frozen or ordered ground state.

Furthermore, by comparing our result with a mostly disordered case [$Cu_{1-\eta}Ga_\eta$]$_A$[$Ga_{2-\eta}Cu_\eta$]$_B O_4$, $\eta$=0.75 [31,50], we can comprehend how the magnetic ions in the *B*-site work. As magnetic ions present in the *B*-site outnumbers those in the *A*-site, the magnetic interactions between the *A*- and *B*-sites (*J*), and the *B*-sites (*K*) play a dominant role in determining the ground state. For instance, G. A. Petrakovskii *et al*. [31] assumed $CuO_6$ octahedra in their sample as being tetragonally distorted in random directions due to Jahn-Teller distortion. They explained its glassy nature using a model with antiferromagnetic *J*=12 K (with modulation $\Delta(J)$=1.2 K due to the distortion) and *K*=6 K. Moreover, below *T*\*=2.5 K, the local magnetism probed using $\mu$SR method captures electronic spins ($\Delta_e \sim$8 MHz) fluctuating at the rate $\nu$=3.7 MHz. This feature is comparable to the term in $CuAl_2O_4$, depicting spins fluctuating in a relatively slower rate $\nu$=7(1) $\mu s^{-1}$ (Fig. A1). It implies that site-disorder is the reason why it can induce the observed glass-like properties in $CuAl_2O_4$. However, $\Delta_e$=151(66) MHz is much larger than that of $CuGa_2O_4$ because it is due to the low concentration of the *B*-site in $CuAl_2O_4$, which leads to a broader distribution of local fields. Under the high longitudinal field of 3 kG, where the second term of Eq. (3) dominates, the partially decoupled shape of $CuAl_2O_4$ depolarization spectrum (solid line 6 in Fig. 5b) originates from the large $\Delta_e$. Due to the large $\nu \approx$115 $\mu s^{-1}$, moreover, the depolarization curve under the low field of 30 G (solid line 5 in Fig. 5b), where the first term of Eq. (3) prevails, does not have an upturn found in $CuGa_2O_4$ (dotted lines 3). It shows instead that the dynamic term in $CuAl_2O_4$ behaves differently from the relaxation of the glass phase in $CuGa_2O_4$. All these considerations given above reinforce our view that the dynamic contribution of the ground state of $CuAl_2O_4$ arises from the possible liquid state of the frustrated diamond lattice, and the glass-like contribution emerges from the site-disorder.

## V. CONCLUSIONS

We have studied $CuAl_2O_4$, a material which realizes the quantum limit of the frustrated *A*-site spinel system with the finite disorder ($\eta$=0.3). To observe the nature of the ground state, we used two local magnetic probes, nuclear magnetic resonance (NMR) and muon spin relaxation ($\mu$SR) methods. The absence of peak splitting in NMR spectra and the non-oscillating feature in $\mu$SR spectra verify a ground state with only short-ranged order below the correlation temperature *T*\*. Instead of completely freezing, the ground state shows the dynamic spin fluctuations characterized by the Gaussian decay in the $\mu$SR spectra.

**ACKNOWLEDGMENTS**

We acknowledge the useful discussions with D. Khomskii, R. Coldea, and C. H. Kim. The work at the IBS CCES was supported by the Institute for Basic Science in Korea (IBS-R009-G1). H.B. acknowledges funding from the European Research Council (ERC) under the European Union's Horizon's 2020 research and innovation program Grant Agreement Number 788814. Part of this work was performed at the Science and Technology Facilities Council (STFC) ISIS Facility, Rutherford Appleton Laboratory, UK. R.N. thanks STFC, Rutherford Appleton Laboratory for beamtime (Experiment RB number 1768046) and DST, India nanomission program for travel support. J.G.P. was partially supported by the Leading Researcher Program of the National Research Foundation of Korea (Grant No. 2020R1A3B2079375).**APPENDIX: The field dependence of $\lambda_2'$ in Eq. (3)**

The field dependence of $\lambda_2'$ in Eq. (3) is given in Fig. A1. To estimate the distribution of local fields and fluctuation frequency, firstly, we used the relation $\lambda_2' = 2\Delta_e^2 \nu/(\nu^2 + \gamma_\mu^2(\mu_0 H_{LF})^2)$ to fit the data. We can roughly fit the data (dashed line) with $\Delta_e$=3.0(2), $\nu$=14(2) μs$^{-1}$, but this fails for fields above 0.5 kG. This relation is based on the simple exponential form for the spin dynamic autocorrelation function. Instead of the simple exponential form, we can use a more general form of the autocorrelation function [51], and this leads to a relaxation rate with which we can fit all the data (solid line) with $\Delta_e$=151(66) and $\nu$=7(1) μs$^{-1}$. The smaller value of fluctuation frequency $\nu$ comparing with that of the sporadic fluctuation term suggests that the local field corresponding to the second term of Eq. (3) has slower dynamics, more comparable to the value $\nu$=3.7 MHz found for $CuGa_2O_4$ [31].

**Figures and Table**

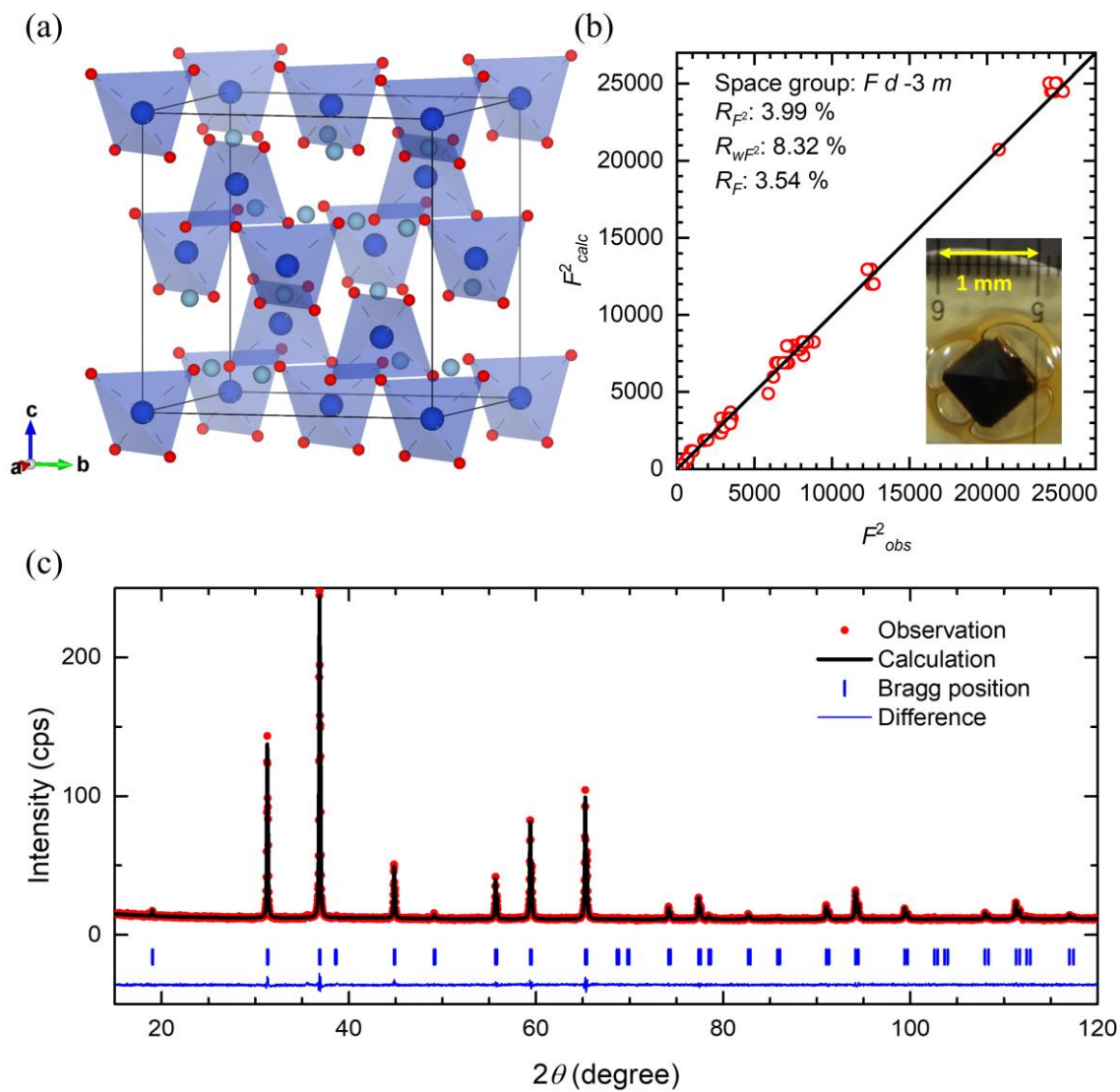

FIG. 1. (a) The crystal structure of a spinel $AB_2X_4$. The blue, grey, and red spheres represent $A$-, $B$-, and $X$-sites, and, in the case of $CuAl_2O_4$, each site is occupied by $Cu^{2+}$, $Al^{3+}$, and $O^{2-}$ ions, respectively. The $CuO_4$ tetrahedra form a diamond lattice. (b) The refinement result of the single-crystal XRD data of $CuAl_2O_4$ and (inset) a picture of a $CuAl_2O_4$ single-crystal. (c) The refinement result of the powder XRD data.

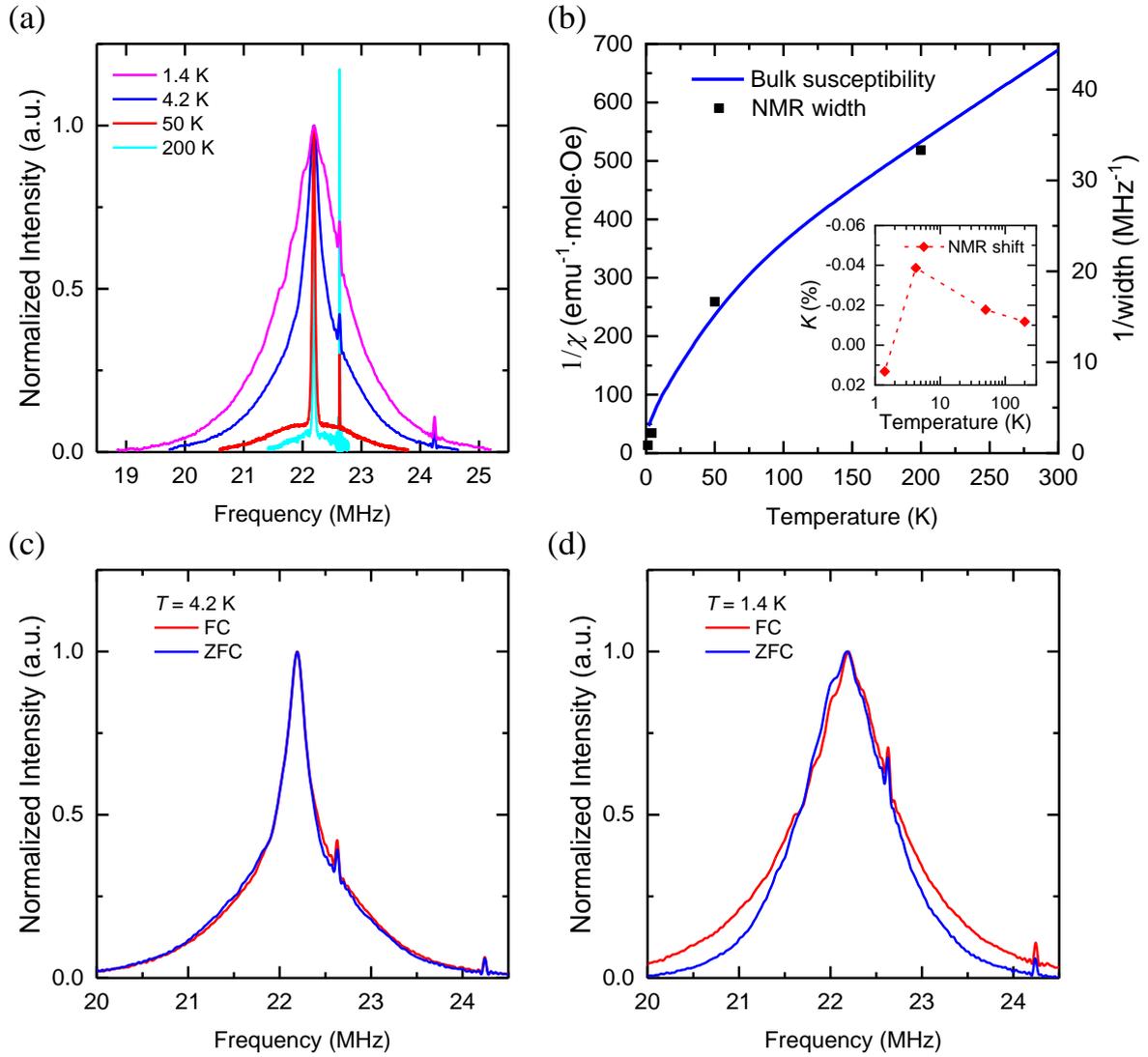

FIG. 2. $^{27}$Al NMR spectra for a single-crystal $CuAl_2O_4$ measured under the external magnetic field $B_0$=20 kG parallel to the crystallographic [0 0 1] direction. (a) The temperature dependence of NMR spectra collected at 1.4, 4.2, 50, and 200 K. The sharp two peaks at 22.630 and 24.243 MHz originate from $^{63}$Cu and $^{65}$Cu nuclei in the NMR coil. (b) Comparison between the inverse of the bulk magnetic susceptibility and the FWHM of the NMR spectra. (Inset) The temperature dependence of the shift of the NMR central peak. (c) and (d) The field-cooled (FC) and zero-field-cooled (ZFC) spectra comparison below and above $T^*$, respectively. The small two satellite peaks come from the coil.

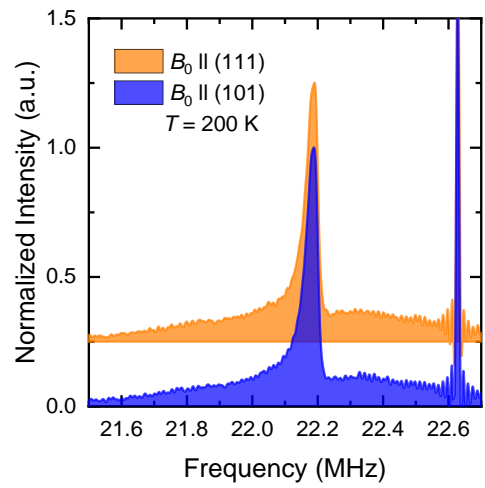

FIG. 3. NMR spectra for the magnetic fields parallel to the crystallographic [1 0 1] and [1 1 1] directions.

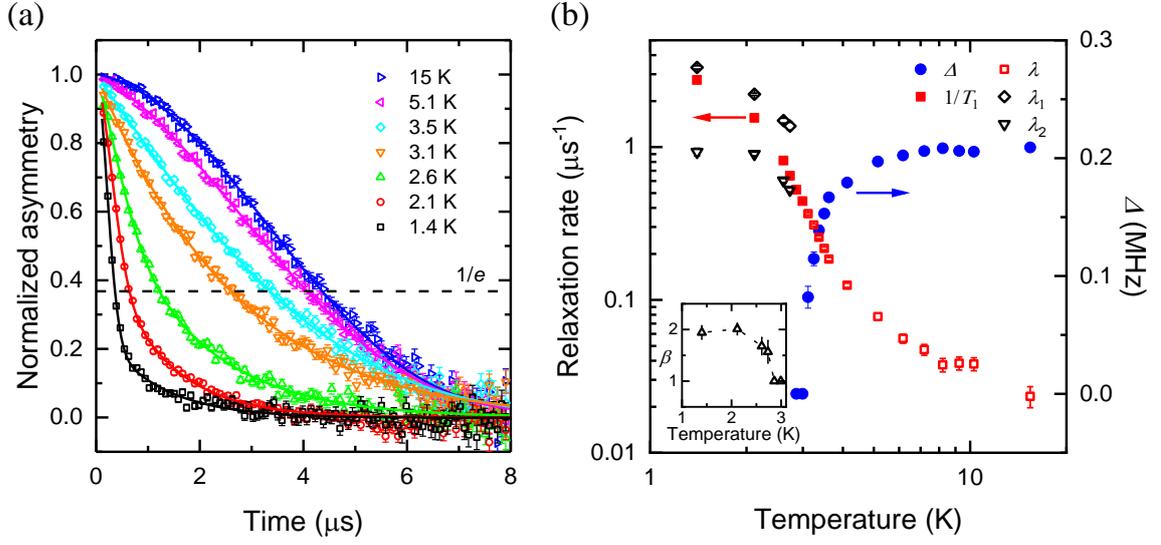

FIG. 4. (a) Zero-field $\mu$SR spectra measured down to 1.4 K. The dashed line indicates $1/e$ decay from the initial polarization. The depolarization spectra are fitted to decay functions Eq. (1) or (2) depicted in the solid lines. (b) (Left) The temperature dependence of relaxation rates obtained from measuring the relaxation time to reach $1/e$ (solid square) and fitting to the relaxation functions: $\lambda$ (hollowed square), $\lambda_1$ (diamond), and $\lambda_2$ (inverted triangle). (Right) The width of the random field distribution $\Delta$, generated by the nuclear moment (circle). (Inset) The temperature dependence of $\beta$ in Eq. (2).

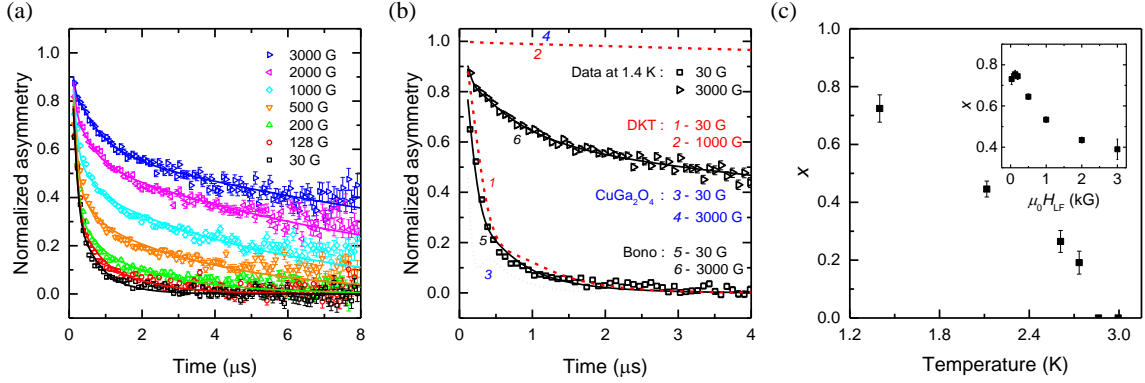

FIG. 5. (a) The longitudinal magnetic field ($H_{LF}$) dependence of the muon depolarization at 1.4 K. The spectra are fitted to a relaxation function (solid line) in Eq. (3). (b) The relaxation spectra at 1.4 K under $\mu_0 H_{LF}$=30 and 3000 G, and calculations using different types of relaxation functions: dynamic Kubo-Toyabe (DKT) function (dashed lines 1 and 2) and Eq. (3) (solid lines 5 and 6). The relaxation of $CuGa_2O_4$ is simulated using the DKT function (dotted lines 3 and 4). (c) The temperature and (inset) longitudinal field dependence of $x$ evaluated from Eq. (2) and Eq. (3), respectively.

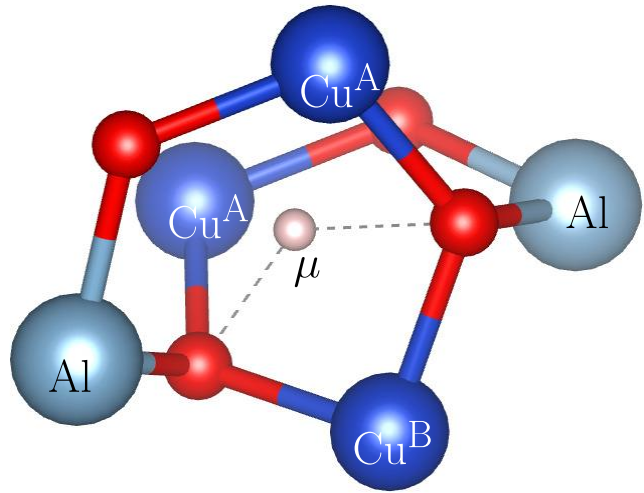

FIG. 6. The local environment around the muon based on a DFT calculation of the electrostatic potential and including site mixing. The muon lies between two oxygen anions (its nearest neighbors) but will be sensitive to fluctuations on three copper cations (its next-nearest neighbors), two on the usual *A*-site, and one on the site-mixed *B*-site.

TABLE 1. The structural information of CuAl$_2$O$_4$ obtained from the refinement shown in Fig. 1.

Single-crystal XRD refinement

| Atom | Wyckoff letter | Site symmetry | x/a | y/b | z/c | $B_{iso}$ (Å$^2$) | Occupancy (%) |
|---|---|---|---|---|---|---|---|
| \multicolumn{8}{c}{Space group: $Fd\bar{3}m$ (No. 227)} |
| \multicolumn{8}{c}{Cell dimensions: a(Å) = 8.083(5), V(Å$^3$) = 528.1(5)} |
| Cu | a | $\bar{4}3m$ | 0 | 0 | 0 | 0.613(52) | 71(2) |
| Al | a | $\bar{4}3m$ | 0 | 0 | 0 | 0.613(52) | 29(2) |
| Al | d | $\bar{3}m$ | 0.625 | 0.625 | 0.625 | 0.267(54) | 86(1) |
| Cu | d | $\bar{3}m$ | 0.625 | 0.625 | 0.625 | 0.267(54) | 14(1) |
| O | e | $3m$ | 0.3860(2) | 0.3860(2) | 0.3860(2) | 0.913(64) | 100 |

Agreement factors: $R_{F2}$(%) = 3.99, $R_{wF2}$(%) = 8.32, $R_F$(%) = 3.54

Powder XRD refinement

| Atom | Wyckoff letter | Site symmetry | x/a | y/b | z/c | $B_{iso}$ (Å$^2$) | Occupancy (%) |
|---|---|---|---|---|---|---|---|
| \multicolumn{8}{c}{Space group: $Fd\bar{3}m$ (No. 227)} |
| \multicolumn{8}{c}{Cell dimensions: a(Å) = 8.081(1), V(Å$^3$) = 527.7(1)} |
| Cu | a | $\bar{4}3m$ | 0 | 0 | 0 | 0.619(252) | 68(5) |
| Al | a | $\bar{4}3m$ | 0 | 0 | 0 | 0.619(252) | 32(5) |
| Al | d | $\bar{3}m$ | 0.625 | 0.625 | 0.625 | 0.498(283) | 84(2) |
| Cu | d | $\bar{3}m$ | 0.625 | 0.625 | 0.625 | 0.498(283) | 16(2) |
| O | e | $3m$ | 0.3863(13) | 0.3863(13) | 0.3863(13) | 0.985(434) | 100 |

Agreement factors: $R_{Bragg}$(%) = 5.94, $R_F$(%) = 6.92

**APPENDIX Figure**

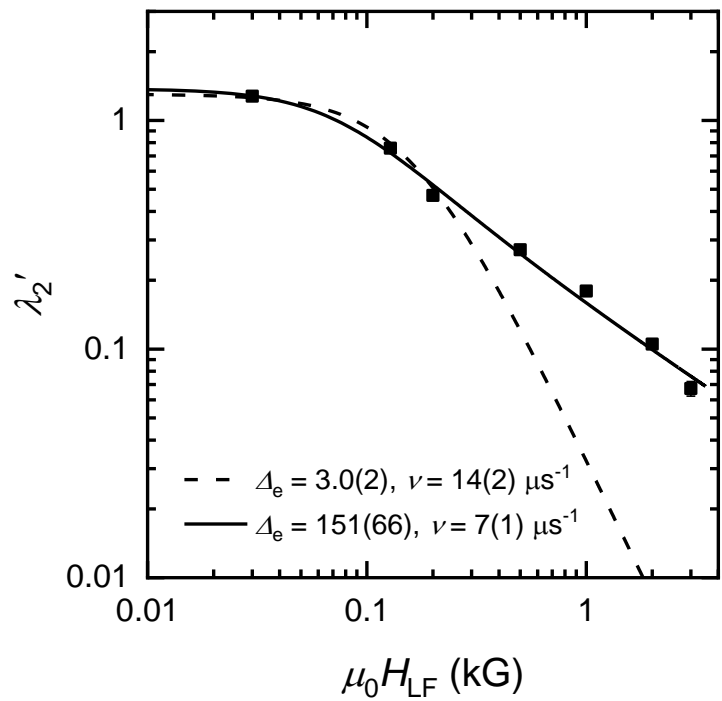

FIG. A1. The external magnetic field dependence of $\lambda_2'$.